\documentclass[submission,copyright,creativecommons]{eptcs}
\providecommand{\event}{ACL2 2017} 
\usepackage{breakurl}             
\usepackage{underscore}           

\title{The Cayley-Dickson Construction in ACL2}
\author{John Cowles \qquad Ruben Gamboa
\institute{
  University of Wyoming \\
  Laramie, WY
}
\email{\{cowles,ruben\}@uwyo.edu}
}
\date{12 January 2017}

\def\titlerunning{Cayley-Dickson in ACL2}
\def\authorrunning{J. Cowles \&\ R. Gamboa}
\begin{document}
\maketitle

\begin{abstract}
 The \textbf{Cayley-Dickson Construction} is a generalization of the
 familiar construction of the complex numbers from pairs of real
 numbers. The complex numbers can be viewed as two-dimensional
 vectors equipped with a multiplication.

 The construction can be used to construct, not only the two-dimensional \textsf{Complex Numbers}, but also the four-dimensional
 \textsf{Quaternions} and the eight-dimensional \textsf{Octonions}.
 Each of these vector spaces has a vector multiplication,
 $\mathbf{v}_{1} \bullet \mathbf{v}_{2}$, that satisfies:
 \begin{enumerate}
   \item Each nonzero vector, $\mathbf{v}$, has a multiplicative inverse
         $\mathbf{v}^{-1}$.
   \item For the Euclidean length of a vector $|\mathbf{v}|$,
    {\boldmath
     $|\mathbf{v}_{1} \bullet \mathbf{v}_{2}| = |\mathbf{v}_{1}| \cdot |\mathbf{v}_{2}|$}
 \end{enumerate}
 \textsf{Real numbers} can also be viewed as (one-dimensional)
 vectors with the above two properties.

 ACL2(r) is used to explore this question: Given a vector space,
 equipped with a multiplication, satisfying the Euclidean length
 condition 2, given above. Make pairs of vectors into ``new''
 vectors with a multiplication.  When do the newly constructed
 vectors also satisfy condition 2?

\end{abstract}

\section{Cayley-Dickson Construction}

Given a vector space, with vector addition, {\boldmath $\mathbf{v}_{1} + \mathbf{v}_{2}$};
vector minus {\boldmath $-\mathbf{v}$}; a zero vector {\boldmath $\vec{0}$};
scalar multiplication by real number $a$, $a \circ \mathbf{v}$;
a unit vector {\boldmath $\vec{1}$}; and vector multiplication 
$\mathbf{v}_{1} \bullet \mathbf{v}_{2}$; satisfying the Euclidean length
condition {\boldmath
     $|\mathbf{v}_{1} \bullet \mathbf{v}_{2}| = |\mathbf{v}_{1}| \cdot |\mathbf{v}_{2}|$} (2).

Define the \textbf{norm} of vector $\mathbf{v}$ by 
{\boldmath $ \|\mathbf{v} \| = |\mathbf{v}|^{2}$}.
Since
{\boldmath $|\mathbf{v}_{1} \bullet \mathbf{v}_{2}| = |\mathbf{v}_{1}| \cdot |\mathbf{v}_{2}|$}
is equivalent to 
{\boldmath
$|\mathbf{v}_{1} \bullet \mathbf{v}_{2}|^{2} = |\mathbf{v}_{1}|^{2} \cdot |\mathbf{v}_{2}|^{2}$},
the Euclidean length condition is equivalent to 
{\boldmath
 \[ \| \mathbf{v}_{1} \bullet \mathbf{v}_{2} \| = \| \mathbf{v}_{1} \| \cdot \| \mathbf{v}_{2} \|. \]}
Recall the \textbf{dot (or inner) product}, of $n$-dimensional vectors,
is defined by 
\begin{eqnarray*}
 (x_{1}, \ldots, x_{n}) \, \mbox{\boldmath $\odot$} \, (y_{1}, \ldots, y_{n})
   & = & x_{1} \cdot y_{1} + \cdots + x_{n} \cdot y_{n} \\
   & = & \sum_{i=1}^{n} x_{i} \cdot y_{i}
\end{eqnarray*}
Then \textsf{Euclidean length} and \textsf{norm} of vector $\mathbf{v}$ are given by
{\boldmath
\[ \begin{array}{ccc}
      | \mathbf{v} |  & = & \sqrt{\mathbf{v} \odot \mathbf{v}} \\
     \| \mathbf{v} \| & = & \mathbf{v} \odot \mathbf{v}.
\end{array} \]}

Except for vector multiplication, it is easy to treat \textsf{ordered pairs of vectors}, 
{\boldmath $(\mathbf{v}_{1} ; \mathbf{v}_{2})$}, as vectors:
\begin{enumerate}
  \item {\boldmath
         $(\mathbf{v}_{1} ; \mathbf{v}_{2}) + (\mathbf{w}_{1} ; \mathbf{w}_{2})
           = (\mathbf{v}_{1} + \mathbf{w}_{1} ; \mathbf{v}_{2} + \mathbf{w}_{2})$}
  \item {\boldmath 
        $-(\mathbf{v}_{1} ; \mathbf{v}_{2}) = (-\mathbf{v}_{1} ; -\mathbf{v}_{2})$}
  \item zero vector: {\boldmath $(\vec{0} ; \vec{0})$}
  \item {\boldmath
         $\mbox{\mathversion{normal} $a$} \circ (\mathbf{v}_{1} ; \mathbf{v}_{2})
           = (\!\! \mbox{\mathversion{normal} $a$} \circ \mathbf{v}_{1} ; 
                \! \mbox{\mathversion{normal} $a$} \circ \mathbf{v}_{2})$}
  \item unit vector: {\boldmath $(\vec{1} ; \vec{0})$}
  \item {\boldmath
         $(\mathbf{v}_{1} ; \mathbf{v}_{2}) \odot (\mathbf{w}_{1} ; \mathbf{w}_{2})
          = \{ \mathbf{v}_{1} \odot \mathbf{w}_{1} \} + \{ \mathbf{v}_{2} \odot \mathbf{w}_{2} \}$}
  \item {\boldmath 
         $\|(\mathbf{v}_{1} ; \mathbf{v}_{2})\| = \|\mathbf{v}_{1}\| + \|\mathbf{v}_{2}\|$}
\end{enumerate}
Given that
{\boldmath
 \[ \| \mathbf{v} \bullet \mathbf{w} \| = \| \mathbf{v} \| \cdot \| \mathbf{w} \|, \]}
the problem is to define \textbf{multiplication of vector pairs} so that
{\boldmath
\[ \|(\mathbf{v}_{1} ; \mathbf{v}_{2}) \bullet (\mathbf{w}_{1} ; \mathbf{w}_{2}) \|
  = \| (\mathbf{v}_{1} ; \mathbf{v}_{2}) \| \cdot \| (\mathbf{w}_{1} ; \mathbf{w}_{2}) \|. \]}

\subsection{Examples}
\begin{description}
 \item[Complex multiplication.] Think of the real numbers as one-dimensional vectors.
     Interpret ordered pairs of real numbers as complex numbers: For real $a$ and $b$,
     $(a ; b) = (\mbox{\textsf{complex}} \ a \ \ b) = a + b \cdot i$.
     For reals $a_{1}, a_{2}$ and $b_{1}, b_{2}$, complex multiplication is defined by
     \begin{equation}
  (a_{1} ; a_{2}) \cdot (b_{1} ; b_{2}) = ([a_{1} b_{1} - a_{2} b_{2}] \, ; \, [a_{1} b_{2} + a_{2} b_{1}])
                              \label{eq:cm}
     \end{equation}
     and satisfies
 \[ \|(a_{1} ; a_{2}) \cdot (b_{1} ; b_{2}) \| = \|(a_{1} ; a_{2}) \| \cdot \| (b_{1} ; b_{2}) \|. \]
   \item[Quaternion multiplication.] Think of the complex numbers as two-dimensional vectors.
     Interpret ordered pairs of complex numbers as William Hamilton's
     quaternions~\cite{conway,koecher,quat}:
     For complex $c = a_{1} + a_{2} \cdot i$ and $d = b_{1} + b_{2} \cdot i$,
     \begin{eqnarray*}
      (c ; d) & = & c + d \cdot j \\
              & = & a_{1} + a_{2} \cdot i + b_{1} \cdot j + b_{2} \cdot i j \\
              & = & a_{1} + a_{2} \cdot i + b_{1} \cdot j + b_{2} \cdot k.
     \end{eqnarray*}
     where $i \cdot j = k$.

     For complex $c = a_{1} + a_{2} \cdot i$, $\bar{c} = a_{1} - a_{2} \cdot i$ is the 
     \textbf{conjugate} of $c$. 

     For complex numbers $c_{1} = a_{1} + a_{2} \cdot i$, $c_{2} = a_{3} + a_{4} \cdot i$,
     $d_{1} = b_{1} + b_{2} \cdot i$, and $d_{2} = b_{3} + b_{4} \cdot i$, quaternion multiplication is 
     defined by
     \begin{eqnarray}
(c_{1} ; c_{2}) \cdot (d_{1} ; d_{2})
                           & = &           (a_{1}b_{1} - a_{2}b_{2} - a_{3}b_{3} - a_{4}b_{4})
                                                                                   \nonumber \\
                           &   & \mbox{} + (a_{1}b_{2} + a_{2}b_{1} + a_{3}b_{4} - a_{4}b_{3}) i 
                                                                                   \nonumber \\
                           &   & \mbox{} + (a_{1}b_{3} - a_{2}b_{4} + a_{3}b_{1} + a_{4}b_{2}) j 
                                                                                   \nonumber \\
                           &   & \mbox{} + (a_{1}b_{4} + a_{2}b_{3} - a_{3}b_{2} + a_{4}b_{1}) k
                                                                                   \nonumber \\
                           & = &           (a_{1}b_{1}- a_{2}b_{2}) + (a_{1}b_{2} + a_{2}b_{1}) i
                                                                                   \nonumber \\
                           &   & \mbox{} - [(a_{3}b_{3} + a_{4}b_{4}) + (-a_{3}b_{4} + a_{4}b_{3}) i]
                                                                                   \nonumber  \\
                           &   & \mbox{} + [(a_{1}b_{3} - a_{2}b_{4}) + (a_{1}b_{4} + a_{2}b_{3}) i] j
                                                                                   \nonumber  \\
                           &   & \mbox{} + [(a_{3}b_{1} + a_{4}b_{2}) + (-a_{3}b_{2} + a_{4}b_{1}) i] j
                                                                                   \nonumber \\
                           & = & (c_{1}d_{1} - c_{2}\bar{d}_{2}) + (c_{1}d_{2} + c_{2}\bar{d}_{1}) j
                                                                                   \nonumber \\
                           & = & ([c_{1}d_{1} - c_{2}\bar{d}_{2}] ; [c_{1}d_{2} + c_{2}\bar{d}_{1}])
                                                     \label{eq:qm}
     \end{eqnarray}
     and satisfies
 \[ \|(c_{1} ; c_{2}) \cdot (d_{1} ; d_{2}) \| = \|(c_{1} ; c_{2}) \| \cdot \| (d_{1} ; d_{2}) \|. \]

     Quaternion multiplication is completely determined by this table for the multiplication
     of $i$, $j$, and $k$:
      \[ \begin{array}{c|rrr}
                 &  i &  j &  k \\
                 \hline
               i & -1 &  k & -j \\
               j & -k & -1 &  i \\
               k &  j & -i & -1
       \end{array} \]
    Quaternion multiplication is \textsf{not} commutative, since $i \cdot j = k$ and
    $j \cdot i = -k$. Quaternion multiplication is associative:
    $(q_{1} \cdot q_{2}) \cdot q_{3} = q_{1} \cdot (q_{2} \cdot q_{3})$. 
   \item[Octonion multiplication.] Think of the quaternions as four-dimensional vectors.
     Interpret ordered pairs of quaternions as John Graves's and Arthur Cayley's
     octonions~\cite{conway,koecher,quat}:
     For quaternions $q = a_{1} + a_{2} \cdot i + a_{3} \cdot j + a_{4} \cdot k$ and 
     $r = b_{1} + b_{2} \cdot i + b_{3} \cdot j + b_{4} \cdot k$,
     \begin{eqnarray*}
      (q ; r)
         & = & q + r \cdot \ell \\
         & = & a_{1} + a_{2} \cdot i + a_{3} \cdot j + a_{4} \cdot k \\
         &   & \mbox{} + b_{1} \cdot \ell + b_{2} \cdot i\ell + b_{3} \cdot j\ell + b_{4} \cdot k\ell \\
         & = &  a_{1} + a_{2} \cdot i + a_{3} \cdot j + a_{4} \cdot k \\
         &   & \mbox{} + b_{1} \cdot \ell + b_{2} \cdot I + b_{3} \cdot J + b_{4} \cdot K.
     \end{eqnarray*}
     where $i \cdot \ell = I$, $j \cdot \ell = J$, and $k \cdot \ell = K$

     For quaternion $q = a_{1} + a_{2} \cdot i + a_{3} \cdot j + a_{4} \cdot k$,
     $\bar{q} = a_{1} - a_{2} \cdot i - a_{3} \cdot j - a_{4} \cdot k$ 
     is the \textbf{conjugate} of $q$.

     For quaternions $q_{1}$, $q_{2}$, $r_{1}$, and $r_{2}$, octonion multiplication is defined by 
     \begin{eqnarray}
     (q_{1} ; q_{2}) \cdot (r_{1} ; r_{2}) 
                          & = & (q_{1} + q_{2} \cdot \ell) \cdot (r_{1} + r_{2} \cdot \ell) \nonumber \\
                          & = & (q_{1} r_{1} - \bar{r}_{2} q_{2})
                                     + (r_{2} q_{1}  + q_{2} \bar{r}_{1}) \cdot \ell \nonumber \\
                          & = & ([q_{1} r_{1} - \bar{r}_{2} q_{2}] ; [r_{2} q_{1}  + q_{2} \bar{r}_{1}])
                                         \label{eq:om}
     \end{eqnarray}
     and satisfies
 \[ \|(q_{1} ; q_{2}) \cdot (r_{1} ; r_{2}) \| = \|(q_{1} ; q_{2}) \| \cdot \| (r_{1} ; r_{2}) \|. \]

     Octonion multiplication is completely determined by this table for the multiplication
     of $i$, $j$, $k$, $\ell$, $I$, $J$, and $K$:
     \[ \begin{array}{c|rrrrrrr}
                &  i   &  j    &  k   &  \ell &  I    &  J    &  K   \\
                \hline  
              i & -1   &  k    & -j   &  I    & -\ell & -K    &  J    \\ 
              j & -k   & -1    &  i   &  J    &  K    & -\ell & -I    \\ 
              k &  j   & -i    & -1   &  K    & -J    &  I    & -\ell \\                              
           \ell & -I   & -J    & -K   & -1    &  i    &  j    &  k    \\           
              I & \ell & -K    &  J   & -i    & -1    & -k    &  j    \\                
              J &  K   &  \ell & -I   & -j    &  k    & -1    & -i    \\ 
              K & -J   &  I    & \ell & -k    & -j    &  i    & -1
     \end{array} \]
    Since the octonions contain the quaternions, octonion multiplication is also \textsf{not} 
    commutative. Octonian multiplication is \textsf{not} associative, since
    $\ell \cdot (I \cdot J) = K$ and $(\ell \cdot I) \cdot J = -K$.
\end{description}
Complex~(\ref{eq:cm}), quaternion~(\ref{eq:qm}), and octonion~(\ref{eq:om}) multiplication 
suggest these possible definitions for vector multiplication:
{\boldmath
\begin{eqnarray}
  (\mathbf{v}_{1} ; \mathbf{v}_{2}) \bullet (\mathbf{w}_{1} ; \mathbf{w}_{2})
   & = & ([\mathbf{v}_{1}\mathbf{w}_{1} - \mathbf{v}_{2}\mathbf{w}_{2}]
            ; [\mathbf{v}_{1}\mathbf{w}_{2} + \mathbf{v}_{2}\mathbf{w}_{1}]) \label{eq1:cm} \\
  (\mathbf{v}_{1} ; \mathbf{v}_{2}) \bullet (\mathbf{w}_{1} ; \mathbf{w}_{2})
   & = & ([\mathbf{v}_{1}\mathbf{w}_{1} - \mathbf{v}_{2}\bar{\mathbf{w}}_{2}]
            ; [\mathbf{v}_{1}\mathbf{w}_{2} + \mathbf{v}_{2}\bar{\mathbf{w}}_{1}]) \label{eq1:qm} \\
  (\mathbf{v}_{1} ; \mathbf{v}_{2}) \bullet (\mathbf{w}_{1} ; \mathbf{w}_{2})
   & = & ([\mathbf{v}_{1}\mathbf{w}_{1} - \bar{\mathbf{w}}_{2}\mathbf{v}_{2}]
            ; [\mathbf{w}_{2}\mathbf{v}_{1} + \mathbf{v}_{2}\bar{\mathbf{w}}_{1}]) \label{eq1:om}
 \end{eqnarray}}
When the $\mathbf{v}_{i}$ and $\mathbf{w}_{i}$ are \textsf{real numbers}, all three definitions are
equivalent, since real multiplication is commutative and the conjugate of real $a$,
$\bar{a}$, is just $a$. 

When the $\mathbf{v}_{i}$ and $\mathbf{w}_{i}$ are \textsf{complex numbers}, 
the last two definitions are equivalent, since complex multiplication is commutative.
However, when $\mathbf{v}_{1} = 1 + 0 \cdot i = \mathbf{w}_{1}$, $\mathbf{v}_{2} = 0 + 1 \cdot i$, 
and $\mathbf{w}_{2} = 0 + (-1 \cdot i)$, the product given by equation~(\ref{eq1:cm}) is the
zero vector $([0 + 0 \cdot i] ; [0 + 0 \cdot i])$. So a vector product defined by (\ref{eq1:cm})
need not satisfy
{\boldmath
\[ \|(\mathbf{v}_{1} ; \mathbf{v}_{2}) \bullet (\mathbf{w}_{1} ; \mathbf{w}_{2}) \|
  = \| (\mathbf{v}_{1} ; \mathbf{v}_{2}) \| \cdot \| (\mathbf{w}_{1} ; \mathbf{w}_{2}) \|, \]}
for complex inputs.

Since the quaternions contain the complex numbers, a vector product defined by (\ref{eq1:cm})
also will not be satisfactory for quaternion inputs.
When
$\mathbf{v}_{1} = 0 + (-1 \cdot i) + 0 \cdot j + 0 \cdot k$,
$\mathbf{v}_{2} = 0 + 0 \cdot i + 1 \cdot j + 0 \cdot k = \mathbf{w}_{2}$, and
$\mathbf{w}_{1} = 0 + 1 \cdot i + 0 \cdot j + 0 \cdot k$,
the product given by equation~(\ref{eq1:qm}) is the zero vector 
$([0 + 0 \cdot i + 0 \cdot j + 0 \cdot k] ; [0 + 0 \cdot i + 0 \cdot j + 0 \cdot k])$.
So a vector product defined by either (\ref{eq1:cm}) or (\ref{eq1:qm})
need not satisfy
{\boldmath
\[ \|(\mathbf{v}_{1} ; \mathbf{v}_{2}) \bullet (\mathbf{w}_{1} ; \mathbf{w}_{2}) \|
  = \| (\mathbf{v}_{1} ; \mathbf{v}_{2}) \| \cdot \| (\mathbf{w}_{1} ; \mathbf{w}_{2}) \|, \]}
for quaternion inputs.

This leaves (\ref{eq1:om}) as a possible way to define vector multiplication of
\textsf{ordered pairs of vectors}. So a vector \textbf{conjugate},
{\boldmath $\bar{\mathbf{v}}$}, is also required. Continuing the enumeration of vector
pair operations given on page 2:
\begin{enumerate}
  \setcounter{enumi}{7}
  \item {\boldmath 
         $\overline{(\mathbf{v}_{1} ; \mathbf{v}_{2})} = (\bar{\mathbf{v}}_{1} ; -\mathbf{v}_{2})$}
  \item {\boldmath
         $(\mathbf{v}_{1} ; \mathbf{v}_{2}) \bullet (\mathbf{w}_{1} ; \mathbf{w}_{2})
           = ([\mathbf{v}_{1}\mathbf{w}_{1} - \bar{\mathbf{w}}_{2}\mathbf{v}_{2}]
            ; [\mathbf{w}_{2}\mathbf{v}_{1} + \mathbf{v}_{2}\bar{\mathbf{w}}_{1}])$}
\end{enumerate}
The enumerated list of items, $1, 2, \cdots, 9$, defining various operations on
\textsf{ordered pairs of vectors}, is called the \textbf{Cayley-Dickson Construction}~\cite{const,quat}.

\subsection{Summary}
\begin{itemize}
   \item Start with the real numbers. \\
         Apply the Cayley-Dickson Construction to pairs of real numbers. \\
         Obtain a vector algebra isomorphic to the complex numbers.
   \item Apply the Cayley-Dickson Construction to pairs of complex numbers. \\
         Obtain a vector algebra isomorphic to the quaternions.
   \item Apply the Cayley-Dickson Construction to pairs of quaternions. \\
         Obtain a vector algebra isomorphic to the octonians.
\end{itemize}
Each of these vector spaces: real numbers, complex numbers, quaternions, and octonians,
satisfy
{\boldmath
 \[ \| \mathbf{v}_{1} \bullet \mathbf{v}_{2} \| = \| \mathbf{v}_{1} \| \cdot \| \mathbf{v}_{2} \|. \]}
Futhermore, every vector {\boldmath $\mathbf{v} \ne \vec{0}$} has a multiplicative inverse:
        {\boldmath
          \[ \mathbf{v}^{-1} = \frac{1}{\| \mathbf{v} \|} \circ \bar{\mathbf{v}} \]}

\section{Composition Algebras}

A \textbf{composition algebra}~\cite{conway,koecher}
is a real vector space, with vector multiplication,
a real-valued \textsf{norm}, and a real-valued \textsf{dot} product,
satisfying this \textsf{composition law}
{\boldmath
\[ \| \mathbf{v}_{1} \bullet \mathbf{v}_{2} \| = \| \mathbf{v}_{1} \| \cdot \| \mathbf{v}_{2} \|. \]}

Use \texttt{encapsulate} to axiomatize, in \textsf{ACL2(r)}, composition algebras. \\
\textsf{ACL2(r)} function symbols are needed for
\begin{itemize}
 \item \textsf{Vector Predicate.} {\boldmath $\mbox{\textbf{Vp}}(x)$}
                                  for ``{\boldmath $x$} is a vector.''
 \item \textsf{Zero Vector.} {\boldmath $\vec{0}$}
 \item \textsf{Vector Addition.} {\boldmath $\mathbf{v}_{1} + \mathbf{v}_{2}$}
 \item \textsf{Vector Minus.} {\boldmath $-\mathbf{v}$}
 \item \textsf{Scalar Multiplication.} For real number $a$, $a \circ \mathbf{v}$
 \item \textsf{Vector Multipication.} $\mathbf{v}_{1} \bullet \mathbf{v}_{2}$
 \item \textsf{Unit Vector.} {\boldmath $\vec{1}$}
 \item \textsf{Vector Norm.} {\boldmath $ \|\mathbf{v} \| $}
 \item \textsf{Vector Dot Product.} {\boldmath $\mathbf{v}_{1} \odot \mathbf{v}_{2}$}
 \item \textsf{Predicate with Quantifier.}
         {\boldmath $ \forall u [\mbox{\textbf{Vp}}(u) \rightarrow u \odot x = 0] $}
 \item \textsf{Skolem Function.} Witness function for Predicate with Quantifier
\end{itemize}
In addition to the usual \textsf{closure axioms}, the \texttt{encapsulate} adds these axioms
to \textsf{ACL2(r)}:
\begin{itemize}
 \item \textsf{Real Vector Space Axioms.}
  {\boldmath
    \[ [\mbox{\textbf{Vp}}(x) \wedge \mbox{\textbf{Vp}}(y) \wedge \mbox{\textbf{Vp}}(z)]
        \rightarrow (x + y) + z = x + (y + z) \]}
  {\boldmath
    \[ [\mbox{\textbf{Vp}}(x) \wedge \mbox{\textbf{Vp}}(y)] \rightarrow x + y = y + x \]}
  {\boldmath
    \[ \mbox{\textbf{Vp}}(x) \rightarrow \vec{0} + x =  x \]}
  {\boldmath
    \[ \mbox{\textbf{Vp}}(x) \rightarrow x + (-x) = \vec{0} \]}
  {\boldmath
    \[ [\mbox{\texttt{Realp}}(a) \wedge \mbox{\texttt{Realp}}(b) \wedge \mbox{\textbf{Vp}}(x)] 
             \rightarrow a \circ (b \circ x) = (a \cdot b) \circ x \]}
  {\boldmath
    \[ \mbox{\textbf{Vp}}(x) \rightarrow 1 \circ x = x \]}
  {\boldmath
    \[ [\mbox{\texttt{Realp}}(a) \wedge \mbox{\texttt{Realp}}(b) \wedge \mbox{\textbf{Vp}}(x)] 
             \rightarrow (a + b) \circ x = (a \circ x) +  (b \circ x) \]}
  {\boldmath
    \[ [\mbox{\texttt{Realp}}(a) \wedge \mbox{\textbf{Vp}}(x) \wedge \mbox{\textbf{Vp}}(y)] 
             \rightarrow a \circ (x + y) = (a \circ x) +  (a \circ y) \]}
 \item \textsf{Real Vector Algebra Axioms.}
  {\boldmath \[ \vec{1} \ne \vec{0} \]}
  {\boldmath
    \begin{eqnarray*}
      \lefteqn{[\mbox{\texttt{Realp}}(a) \wedge \mbox{\texttt{Realp}}(b) \wedge
                \mbox{\textbf{Vp}}(x) \wedge \mbox{\textbf{Vp}}(y) \wedge \mbox{\textbf{Vp}}(z)] 
                \rightarrow}         \\
       & & x \bullet [(a \circ y) + (b \circ z)] =
           [a \circ (x \bullet y)] + [b \circ (x \bullet z)]
     \end{eqnarray*}}
  {\boldmath
    \begin{eqnarray*}
     \lefteqn{[\mbox{\texttt{Realp}}(a) \wedge \mbox{\texttt{Realp}}(b) \wedge
               \mbox{\textbf{Vp}}(x) \wedge \mbox{\textbf{Vp}}(y) \wedge \mbox{\textbf{Vp}}(z)] 
               \rightarrow}                             \\
     & & [(a \circ x) + (b \circ y)] \bullet z  =
         [a \circ (x \bullet z)] + [b \circ (y \bullet z)]
     \end{eqnarray*}}
  {\boldmath
    \[ \mbox{\textbf{Vp}}(x) \rightarrow [(\vec{1} \bullet x =  x) \wedge
                                        (x \bullet \vec{1} =  x)] \]}
 \item \textsf{Vector Norm and Dot Product Axioms.}
   {\boldmath \[\mbox{\textbf{Vp}}(x) \rightarrow
                [\mbox{\texttt{Realp}}(\| x \|) \wedge \| x \| \ge 0] \]}
   {\boldmath \[\mbox{\textbf{Vp}}(x) \rightarrow
                [(\| x \| = 0) = (x = \vec{0})] \]}
   {\boldmath \[ [\mbox{\textbf{Vp}}(x) \wedge \mbox{\textbf{Vp}}(y)] \rightarrow
                \| x \bullet y \| = \| x \| \cdot \| y \| \]}
   {\boldmath \[ x \odot y = \frac{1}{2} \cdot [ \| x + y \| - \| x \| - \| y \| ] \]} 
  {\boldmath
    \begin{eqnarray*}
     \lefteqn{[\mbox{\texttt{Realp}}(a) \wedge \mbox{\texttt{Realp}}(b) \wedge
               \mbox{\textbf{Vp}}(x) \wedge \mbox{\textbf{Vp}}(y) \wedge \mbox{\textbf{Vp}}(z)] 
               \rightarrow}                             \\
     & & [(a \circ x) + (b \circ y)] \odot z  =
         [a \cdot (x \odot z)] + [b \cdot (y \odot z)]
     \end{eqnarray*}}
   {\boldmath
     \begin{eqnarray*}
      \lefteqn{\forall u [\mbox{\textbf{Vp}}(u) \rightarrow u \odot x = 0] = } \\
        & & [\mbox{\texttt{let}} \; u \; \mbox{\texttt{be}} \; \mbox{\texttt{witness}}(x)]
               [\mbox{\textbf{Vp}}(u) \rightarrow u \odot x = 0]
      \end{eqnarray*}}
   {\boldmath
     \[ \forall u [\mbox{\textbf{Vp}}(u) \rightarrow u \odot x = 0] \rightarrow
               [\mbox{\textbf{Vp}}(u) \rightarrow u \odot x = 0] \]}
   {\boldmath \[(\mbox{\textbf{Vp}}(x) \wedge
                 \forall u [\mbox{\textbf{Vp}}(u) \rightarrow u \odot x = 0])
               \rightarrow
               x = \vec{0}\]}
\end{itemize}
The \textsf{ACL2(r)} theory of \textsf{composition algebras} includes the following theorems
and definitions:
\begin{itemize}
  \item \textsf{Scaling Laws.}
    {\boldmath
      \[ [\mbox{\textbf{Vp}}(x) \wedge \mbox{\textbf{Vp}}(y) \wedge \mbox{\textbf{Vp}}(z)]
        \rightarrow  (x \bullet y) \odot (x \bullet z) = \| x \| \cdot (y \odot z) \]}
    {\boldmath
      \[ [\mbox{\textbf{Vp}}(x) \wedge \mbox{\textbf{Vp}}(y) \wedge \mbox{\textbf{Vp}}(z)]
        \rightarrow  (x \bullet z) \odot (y \bullet z) = (x \odot y) \cdot  \| z \| \]}
  \item \textsf{Exchange Law.}
    {\boldmath
       \begin{eqnarray*}
         \lefteqn{[\mbox{\textbf{Vp}}(u) \wedge \mbox{\textbf{Vp}}(x) \wedge 
           \mbox{\textbf{Vp}}(y) \wedge \mbox{\textbf{Vp}}(z)] \rightarrow} \\
             & &  [(u \bullet y) \odot (x \bullet z)] +
                  [(u \bullet z) \odot (x \bullet y)] =         \\
             & &  2 \cdot (u \odot x) \cdot (y \odot z) 
        \end{eqnarray*}}
  \item \textsf{Conjugate Definition.}
     {\boldmath
       \[ \bar{x} = ([2 \cdot (x \odot \vec{1})] \circ \vec{1}) + (-x) \]} 
  \item \textsf{Conjugate Laws.}
    {\boldmath
      \[ [\mbox{\textbf{Vp}}(x) \wedge \mbox{\textbf{Vp}}(y) \wedge \mbox{\textbf{Vp}}(z)]
        \rightarrow y \odot (\bar{x} \bullet z) = z \odot (x \bullet y) \]}  
    {\boldmath
      \[ [\mbox{\textbf{Vp}}(x) \wedge \mbox{\textbf{Vp}}(y) \wedge \mbox{\textbf{Vp}}(z)]
        \rightarrow x \odot (z \bullet \bar{y}) = z \odot (x \bullet y) \]}  
  {\boldmath
    \[ \mbox{\textbf{Vp}}(x) \rightarrow \bar{\bar{x}} = x \]}
  {\boldmath
    \[ [ \mbox{\textbf{Vp}}(x) \wedge \mbox{\textbf{Vp}}(y)] 
             \rightarrow \overline{x \bullet y} = \bar{y} \bullet \bar{x} \]}
  \item \textsf{Inverse Definition.}
     {\boldmath
          \[ x^{-1} = \frac{1}{\| x \|} \circ \bar{x} \]}
  \item \textsf{Inverse Law.}
     {\boldmath
      \[ [\mbox{\textbf{Vp}}(x) \wedge x \ne \vec{0}] \rightarrow
         [ x^{-1} \bullet x = \vec{1} \wedge x \bullet x^{-1} = \vec{1} \, ] \]}
  \item \textsf{Alternative Laws.} Special versions of associativity.
  {\boldmath
    \[ [ \mbox{\textbf{Vp}}(x) \wedge \mbox{\textbf{Vp}}(y)] 
             \rightarrow x \bullet (x \bullet y) = (x \bullet x) \bullet y \]}
  {\boldmath
    \[ [ \mbox{\textbf{Vp}}(x) \wedge \mbox{\textbf{Vp}}(y)] 
             \rightarrow (y \bullet x) \bullet x = y \bullet (x \bullet x) \]}
  {\boldmath
    \[ [ \mbox{\textbf{Vp}}(x) \wedge \mbox{\textbf{Vp}}(y)] 
             \rightarrow (x \bullet y) \bullet x = x \bullet (y \bullet x) \]}
  \item \textsf{Other Theorems.}
   {\boldmath \[ [\mbox{\textbf{Vp}}(x) \wedge \mbox{\textbf{Vp}}(y)] \rightarrow
                 ([x \bullet y = \vec{0}] = [(x = \vec{0}) \vee (y = \vec{0})]) \]} 
   {\boldmath \[(\mbox{\textbf{Vp}}(x) \wedge \mbox{\textbf{Vp}}(y) \wedge
                 \forall u [\mbox{\textbf{Vp}}(u) \rightarrow u \odot x = u \odot y])
               \rightarrow
               x = y \]}
   {\boldmath \[\| x \| = x \odot x\]}
\end{itemize}

\section{Composition Algebra Doubling.}

Use \texttt{encapsulate} to axiomatize, in \textsf{ACL2(r)}, two composition algebras, 
with vector predicates {\boldmath $V_{1}p$} and {\boldmath $V_{2}p$}. The vectors satisfying 
{\boldmath $V_{2}p$} are \texttt{ordered pairs} of elements satisfying {\boldmath $V_{1}p$}.
Both algebras, individually, satisfy all the axioms (and also all theorems) of the previous section.
These additional axioms connect the various vector operations of the two spaces:
\begin{itemize}
 \item \textsf{Additional Axioms.}
   {\boldmath
     \[ x +_{2} y = ([\mbox{\texttt{car}}(x) +_{1} \mbox{\texttt{car}}(y)]
                    ; [\mbox{\texttt{cdr}}(x) +_{1} \mbox{\texttt{cdr}}(y)]) \]}
   {\boldmath
     \[ a \circ_{2} x = ([a \circ_{1} \mbox{\texttt{car}}(x)]
                        ; [a \circ_{1} \mbox{\texttt{cdr}}(x)]) \]}
   {\boldmath
      \[ [V_{1}p(x) \wedge V_{1}p(y)] \rightarrow 
            (x ; \vec{0}_{1}) \bullet_{2} (y ; \vec{0}_{1})
            = ([x \bullet_{1} y] ; \vec{0}_{1}) \]} 
   {\boldmath
      \[ V_{1}p(x) \rightarrow 
           ([ \| (x ; \vec{0}_{1}) \|_{2} = \| x \|_{1} ] \wedge
            [ \| (\vec{0}_{1} ; x) \|_{2} = \| x \|_{1} ]) \]}
   {\boldmath
      \[ [V_{1}p(x) \wedge V_{1}p(y)] \rightarrow 
            (x ; \vec{0}_{1}) \odot_{2} (\vec{0}_{1} ; y) = 0 \]}
   {\boldmath
      \[ \vec{1}_{2} = (\vec{1}_{1} ; \vec{0}_{1}) \]}
   {\boldmath
      \[ V_{1}p(x) \rightarrow 
           (x ; \vec{0}_{1}) \bullet_{2} (\vec{0}_{1} ; \vec{1}_{1})
            = (\vec{0}_{1} ; x) \]}
\end{itemize}
Since both {\boldmath $V_{1}p$} and {\boldmath $V_{2}p$} are composition algebras,
{\boldmath
 \begin{eqnarray*}
  \lefteqn{[V_{1}p(v_{1}) \wedge V_{1}p(v_{2}) \wedge V_{1}p(w_{1}) \wedge V_{1}p(w_{2})]
              \rightarrow} \\
   & & [ \,  \| v_{1} \bullet_{1} v_{2} \|_{1} = \| v_{1} \|_{1} \cdot \| v_{2} \|_{1} \wedge \\
   & & \; \, \|( v_{1} ; w_{1}) \bullet_{2} ( v_{2} ; w_{2}) \|_{2}
             = \| ( v_{1} ; w_{1}) \|_{2} \cdot \| ( v_{2} ; w_{2}) \|_{2} \, ].
 \end{eqnarray*}}
Among the consequences of these \texttt{encapsulated} axioms, \textsf{ACL2(r)} verifies:
\begin {itemize}
 \item \textsf{Dot Product Doubling.}
   {\boldmath
     \begin{eqnarray*}
     \lefteqn{[V_{1}p(v_{1}) \wedge V_{1}p(v_{2}) \wedge V_{1}p(w_{1}) \wedge V_{1}p(w_{2})]
              \rightarrow} \\
      & & ( v_{1} ; v_{2} ) \odot_{2} ( w_{1} ; w_{2} ) 
            = [v_{1} \odot_{1} w_{1}] + [v_{2} \odot_{1} w_{2}]
      \end{eqnarray*}}
 \item \textsf{Conjugation Doubling.}
   {\boldmath
      \[ [V_{1}p(v_{1}) \wedge V_{1}p(v_{2})] \rightarrow
          \overline{( v_{1} ; v_{2})} = (\bar{v}_{1} ; - v_{2}) \]}
 \item \textsf{Product Doubling.}
   {\boldmath
     \begin{eqnarray*}
     \lefteqn{[V_{1}p(v_{1}) \wedge V_{1}p(v_{2}) \wedge V_{1}p(w_{1}) \wedge V_{1}p(w_{2})]
              \rightarrow} \\
      & & ( v_{1} ; v_{2} ) \bullet_{2} ( w_{1} ; w_{2} ) = \\
      & & ( \{[v_{1} \bullet_{1} w_{1}] - [\bar{w}_{2} \bullet_{1} v_{2}] \}
            ; \{ [w_{2} \bullet_{1} v_{1}] +_{1} [v_{2} \bullet_{1} \bar{w}_{1}] \} )
      \end{eqnarray*}}
 \item \textsf{Norm Doubling.}
   {\boldmath
      \[ [V_{1}p(v_{1}) \wedge V_{1}p(v_{2})] \rightarrow
         \|( v_{1} ; v_{2} )\|_{2} = \| v_{1} \|_{1} + \| v_{2} \|_{1} \]}
 \item \textsf{Associativity of} $\bullet_{1}$.
   {\boldmath
      \[ [V_{1}p(v_{1}) \wedge V_{1}p(v_{2}) \wedge V_{1}p(v_{3})] \rightarrow
         [ v_{1} \bullet_{1} v_{2} ] \bullet_{1} v_{3} =
         v_{1} \bullet_{1} [ v_{2} \bullet_{1} v_{3} ] \]}
\end{itemize}
The above \textsf{doubling theorems} match the definitions used in the 
\textsf{Cayley-Dickson Construction} for making \textsf{ordered pairs} of 
{\boldmath $V_{1}p$} vectors into {\boldmath $V_{2}p$} vectors.
Furthermore, if both the {\boldmath $V_{2}p$} \textsf{ordered pairs} and the
the component {\boldmath $V_{1}p$} vectors form composition algebras, then
the component {\boldmath $V_{1}p$} algebra has an associative multiplication.

In addition, \textsf{ACL2(r)} verifies that if the component {\boldmath $V_{1}p$}
vectors form a composition algebra with an associative multiplication, then 
the \textsf{Cayley-Dickson Construction}  makes the {\boldmath $V_{2}p$}
\textsf{ordered pairs} into a composition algebra.

\subsection{Summary}
\textsf{ACL2(r)} verifies:
\begin{itemize}
 \item Start with a composition algebra {\boldmath $V_{1}p$}.
 \item Let {\boldmath $V_{2}p$} be the set of ordered pairs of elements from {\boldmath $V_{1}p$}.
 \item Then 
  \begin{enumerate}
    \item
      \begin{enumerate}
         \item If {\boldmath $V_{2}p$} is also a composition algebra, 
          then {\boldmath $V_{1}p$}-multiplication is \textsf{associative}.
         \item If {\boldmath $V_{1}p$}-multiplication is \textsf{associative}, then
          {\boldmath $V_{2}p$} can be made into a composition algebra.
      \end{enumerate}
    \item
     \begin{enumerate}
       \item If {\boldmath $V_{2}p$} is a composition algebra with
          \textsf{associative} multiplication, then {\boldmath $V_{1}p$}-multiplication
          is \textsf{associative}  and \textsf{commutative}.
       \item If {\boldmath $V_{1}p$}-multiplication is \textsf{associative} and 
              \textsf{commutative}, then {\boldmath $V_{2}p$} can be made into a 
              composition algebra with \textsf{associative} multiplication.
     \end{enumerate}  
    \item
      \begin{enumerate}
        \item If {\boldmath $V_{2}p$} is a composition algebra with \textsf{associative} 
          and \textsf{commutative} multiplication, then {\boldmath $V_{1}p$}-multiplication
          is also \textsf{associative} and \textsf{commutative}, and 
          {\boldmath $V_{1}p$}-conjugation is \textsf{trivial}.
       \item If {\boldmath $V_{1}p$}-multiplication is \textsf{associative} and 
          \textsf{commutative}, and \\ 
             {\boldmath $V_{1}p$}-conjugation is \textsf{trivial}, 
             then {\boldmath $V_{2}p$} can be made into a composition algebra with 
             \textsf{associative} and \textsf{commutative} multiplication.
      \end{enumerate}
   \end{enumerate}
\end{itemize}

\subsubsection{A last example}
Apply the Cayley-Dickson Construction to pairs of octonions.
Think of the octonions as eight-dimensional vectors.
Interpret ordered pairs of octonions as sixteen-dimensional vectors
called \textsf{Sedenions}~\cite{sed}: For octonians
\begin{eqnarray*} 
o & = & a_{1} + a_{2} \cdot i + a_{3} \cdot j + a_{4} \cdot k
        + a_{5} \cdot \ell + a_{6} \cdot I + a_{7} \cdot J + a_{8} \cdot K \\
p & = & b_{1} + b_{2} \cdot i + b_{3} \cdot j + b_{4} \cdot k
        + b_{5} \cdot \ell + b_{6} \cdot I + b_{7} \cdot J + b_{8} \cdot K,
\end{eqnarray*}
\begin{eqnarray*}
 (o ; p)
   & = & o + p \cdot L \\
   & = & a_{1} + a_{2} \cdot i + a_{3} \cdot j + a_{4} \cdot k
       + a_{5} \cdot \ell + a_{6} \cdot I + a_{7} \cdot J + a_{8} \cdot K \\
   &   & \mbox{} + b_{1} \cdot L + b_{2} \cdot i L + b_{3} \cdot j L + b_{4} \cdot k L
       + b_{5} \cdot \ell L + b_{6} \cdot I L \\
   &   & \mbox{} + b_{7} \cdot J L + b_{8} \cdot K L.
\end{eqnarray*}
For octonion
\begin{eqnarray*} 
     o  & = & a_{1} + a_{2} \cdot i + a_{3} \cdot j + a_{4} \cdot k
              + a_{5} \cdot \ell + a_{6} \cdot I + a_{7} \cdot J + a_{8} \cdot K, \\
\bar{o} & = & a_{1} - a_{2} \cdot i - a_{3} \cdot j - a_{4} \cdot k
              - a_{5} \cdot \ell - a_{6} \cdot I - a_{7} \cdot J - a_{8} \cdot K
\end{eqnarray*}
is the \textbf{conjugate} of $o$.

For octonions $o_{1}$, $o_{2}$, $p_{1}$, and $p_{2}$, sedenion multiplication is defined by 
     \begin{eqnarray*}
     (o_{1} ; o_{2}) \cdot (p_{1} ; p_{2}) 
                          & = & (o_{1} + o_{2} \cdot L) \cdot (p_{1} + p_{2} \cdot L) \\
                          & = & (o_{1} p_{1} - \bar{p}_{2} o_{2})
                                     + (p_{2} o_{1}  + o_{2} \bar{p}_{1}) \cdot L  \\
                          & = & ([o_{1} p_{1} - \bar{p}_{2} o_{2}] ; [p_{2} o_{1}  + o_{2} \bar{p}_{1}])
     \end{eqnarray*}
Recall the octonians have a \textbf{non}-trivial conjugate and octonian
multiplication is \textbf{not} commutative and also \textbf{not} associative.
The octonions form a composition algebra, so that for octonions $o$ and $p$,
$ \| o \cdot p \| = \| o \| \cdot \| p \| $.

By item 1(a) listed above about composition algebras, since octonian multiplication is not
associative, the sedenions is not a composition algebra. In fact, the sedenion product of nonzero
vectors could be the zero vector. For example, let $o_{1}$, $o_{2}$, $p_{1}$, and $p_{2}$ be these 
otonions:
\begin{eqnarray*}
o_{1} & = &  0 + 0 \cdot i + 0 \cdot j + 1 \cdot k +
                       0 \cdot \ell + 0 \cdot I + 0 \cdot J + 0 \cdot K \\
o_{2} & = &  0 + 0 \cdot i + 1 \cdot j + 0 \cdot k +
                       0 \cdot \ell + 0 \cdot I + 0 \cdot J + 0 \cdot K \\
p_{1} & = &  0 + 0 \cdot i + 0 \cdot j + 0 \cdot k +
                       0 \cdot \ell + 0 \cdot I + 1 \cdot J + 0 \cdot K \\
p_{2} & = &  0 + 0 \cdot i + 0 \cdot j + 0 \cdot k +
                       0 \cdot \ell + 0 \cdot I + 0 \cdot J + (-1 \cdot K).
\end{eqnarray*}
Then
\begin{eqnarray*}
\lefteqn{(o_{1} + o_{2} \cdot L) \cdot (p_{1} + p_{2} \cdot L) =} \\
      & & (0 + 0 \cdot i + 0 \cdot j + 0 \cdot k +
                       0 \cdot \ell + 0 \cdot I + 0 \cdot J + 0 \cdot K) \\
      & & \mbox{} + (0 + 0 \cdot i + 0 \cdot j + 0 \cdot k +
                       0 \cdot \ell + 0 \cdot I + 0 \cdot J + 0 \cdot K) \cdot L.
\end{eqnarray*}
So \[ \| (o_{1} + o_{2} \cdot L) \cdot (p_{1} + p_{2} \cdot L) \| = 0, \]
but
\[ \| (o_{1} + o_{2} \cdot L) \| = 2 = \| (p_{1} + p_{2} \cdot L) \|, \]
and
\[ \| (o_{1} + o_{2} \cdot L) \cdot (p_{1} + p_{2} \cdot L) \| \ne
   \| (o_{1} + o_{2} \cdot L) \| \cdot \| (p_{1} + p_{2} \cdot L) \|. \]

All nonzero sedenions have multiplicative inverses.
For example,
\begin{eqnarray*}
\lefteqn{(o_{1} + o_{2} \cdot L)^{-1} =} \\
      & & (0 + 0 \cdot i + 0 \cdot j + (-\frac{1}{2} \cdot k) +
                       0 \cdot \ell + 0 \cdot I + 0 \cdot J + 0 \cdot K) \\
      & & \mbox{} + (0 + 0 \cdot i + (-\frac{1}{2} \cdot j) + 0 \cdot k +
                       0 \cdot \ell + 0 \cdot I + 0 \cdot J + 0 \cdot K) \cdot L \\
\lefteqn{(p_{1} + p_{2} \cdot L)^{-1} =} \\
      & & (0 + 0 \cdot i + 0 \cdot j + 0 \cdot k) +
                       0 \cdot \ell + 0 \cdot I + (-\frac{1}{2} \cdot J) + 0 \cdot K) \\
      & & \mbox{} + (0 + 0 \cdot i + 0 \cdot j + 0 \cdot k +
                       0 \cdot \ell + 0 \cdot I + 0 \cdot J + (\frac{1}{2} \cdot K)) \cdot L 
\end{eqnarray*}

\appendix
\section{ACL2(r) Books}
 \subsection{\texttt{cayley1.lisp}}
  The Reals form a (1-dimensional) composition algebra.
 \subsection{\texttt{cayley1a.lisp}}
  Cons pairs of Reals form a (2-dimensional) composition algebra. \\
  This algebra is (isomorphic to) the Complex Numbers.
 \subsection{\texttt{cayley1b.lisp}}
  Cons pairs of Complex Numbers form a (4-dimensional) composition algebra. \\
  This algebra is (isomorphic to) the Quaternions. 

  3-Dimensional Vector Cross Product and 3-Dimensional Dot Product
  are related to 4-Dimensional Quaternion Multiplication.
 \subsection{\texttt{cayley1c.lisp}}
  Cons pairs of Quaternions form a (8-dimensional) composition algebra. \\
  This algebra is (isomorphic to) the Octonions.

  7-Dimensional Vector Cross Product and 7-Dimensional Dot Product
  are related to 8-Dimensional Octonion Multiplication.
 \subsection{\texttt{cayley1d.lisp}}
  Cons pairs of Octonions form a (16-dimensional) algebra. \\
  This algebra is (isomorphic to) the Sedenions.

  This algebra is \textsf{not} a composition algebra, but all
  nonzero Sedenions have multiplicative inverses.
 \subsection{\texttt{cayley2.lisp}}
  Axioms and theory of composition algebras.
 \subsection{\texttt{cayley2.lisp}}
  In composition algebras, $\| v \| = v \odot v$.
 \subsection{\texttt{cayley3.lisp}}
  Start with a composition algebra V1.
  Let V2 be the set of ordered pairs of elements from V1.

  If V2 is also a composition algebra, 
    then V1-multiplication is associative.
 \subsection{\texttt{cayley3a.lisp}}
  Start with a composition algebra V.
  Let V2 be the set of ordered pairs of elements from V.

  If V-multiplication is associative, 
    then V2 can be made into a composition algebra.
 \subsection{\texttt{cayley4.lisp}}
  Start with a composition algebra V1.
  Let V2 be the set of ordered pairs of elements from V1.

  If V2 is a composition algebra with associative multiplication, 
    then V1-multiplication is associative and commutative.
 \subsection{\texttt{cayley4a.lisp}}
  Start with a composition algebra V.
  Let V2 be the set of ordered pairs of elements from V.

  If V-multiplication is associative and commutative, 
    then V2 can be made into a composition algebra with associative multiplication.
 \subsection{\texttt{cayley5.lisp}}
  Start with a composition algebra V1.
  Let V2 be the set of ordered pairs of elements from V1.

  If V2 is a composition algebra with associative and commutative multiplication, 
    then V1-multiplication is associative and commutative, and V1-conjugation is
    trivial.
 \subsection{\texttt{cayley5a.lisp}}
  Start with a composition algebra V.
  Let V2 be the set of ordered pairs of elements from V.

  If V-multiplication is associative and commutative, and V-conjugation is trivial, 
    then V2 can be made into a composition algebra with associative and commutative multiplication.

\nocite{*}
\bibliographystyle{eptcs}
\bibliography{cayley-paper1}

\end{document}